
\tolerance=10000
\input phyzzx

\REF\cst{C.M. Hull, Nucl.Phys. {\bf B583} (2000) 237, 
hep-th/0004195.}
\REF\HT{C.M. Hull and P.K. Townsend, Nucl. Phys. {\bf B438} (1995)
109;  hep-th/9410167.}
\REF\jhs{J. H. Schwarz, Talk presented at the Ninth Marcel Grossmann Meeting
(MG9), hep-th/0008009.}
\REF\Verl{ 
 E. Verlinde,
       Nucl. Phys. {\bf B455} (1995) 211,  hep-th/9506011.}
\REF\notivarg{
S.~Deser, P.~K.~Townsend and W.~Siegel,
Nucl.\ Phys.\  {\bf B184}, 333 (1981).}
\REF\notoph{V.~I.~Ogievetsky and I.~V.~Polubarinov,
Sov.\ J.\ Nucl.\ Phys.\  {\bf 4}, 156 (1967).}
\REF\CJP{E. Cremmer, B. Julia, H. Lu and C.N. Pope, Nucl. Phys. {\bf B535 }
(1998)
242, hep-th/9806106.}
\REF\GZ {M.K. Gaillard and B. Zumino, Nucl. Phys. {\bf B193} (1981) 221.}
\REF\HTE{C.M. Hull and P.K. Townsend, Nucl. Phys. {\bf B451}
(1995)
525,
hep-th/9505073.}
\REF\julia {B. Julia in {\it Supergravity and Superspace}, S.W.
Hawking and M.
Ro\v cek, C.U.P.
Cambridge,  1981. }
\REF\CJ{E. Cremmer and B. Julia, Phys. Lett. {\bf 80B} (1978) 48; Nucl.
Phys. {\bf B159} (1979) 141.}
 \REF\Cremmer{E. Cremmer, in {\it Supergravity and Superspace},
S.W.
Hawking and
M. Ro\v cek, C.U.P.
Cambridge,  1981.}
\REF\intri{
 K. Intriligator and W. Skiba,
      Nucl. Phys. {\bf B559} (1999) 165, hep-th/9905020;
  K. Intriligator, Nucl.Phys. {\bf B551} (1999) 575
  hep-th/9811047.}

%

%
\font\mybb=msbm10 at 12pt
\def\bbcc#1{\hbox{\mybb#1}}
\def\Z {\bbcc{Z}}
\def\R {\bbcc{R}}
\def\C {\bbcc{C}}
\def \aa {\alpha}
\def \bb {\beta}
\def \gg {\gamma}
\def \dd {\delta}
\def \ee {\epsilon}

\def \kk {\kappa}
\def \ll {\lambda}
\def \mm {\mu}
\def \nn {\nu}

\def \rr {\rho}
\def \ss {\sigma}
\def \tt {\tau}

\def \zz {\zeta}
\def \th {\theta}

 \def \ggg {\Gamma}

\def \www{\Omega}

\def\lsemidir{\mathbin{\hbox{\hskip2pt\vrule height 5.7pt depth -.3pt
width .25pt\hskip-2pt$\times$}}}

\def\sd{{\lsemidir}} 
\def \xx {{\sd}}

\def \ti {\tilde}

\def \2 {{1 \over 2}}
\def \3 {{1 \over 3}}
\def \4 {{1 \over 4}}
\def \5 {{1 \over 5}}
\def \6 {{1 \over 6}}
\def \7 {{1 \over 7}}
\def \8 {{1 \over 8}}
\def \9 {{1 \over 9}}
\def \00 { \infty}

\def\++ {{(+)}}
\def \- {{(-)}}
\def\+-{{(\pm)}}

\def \pa {\partial}


 \def\unit{\hbox to 3.3pt{\hskip1.3pt \vrule height 7pt width .4pt
\hskip.7pt
\vrule height 7.85pt width .4pt \kern-2.4pt
\hrulefill \kern-3pt
\raise 4pt\hbox{\char'40}}}

\def\nup#1({Nucl.\ Phys.\  {\bf B#1}\ (}

\def \qq {\qquad}

\def\dalemb#1#2{{\vbox{\hrule height .#2pt
        \hbox{\vrule width.#2pt height#1pt \kern#1 pt
                \vrule width.#2 pt}
        \hrule height.#2 pt}}}
\def\square{\mathord{\dalemb{6.2}{6}\hbox{\hskip1pt}}}
\def\boxx{{\square }}


\Pubnum{ \vbox{ \hbox {QMW-PH/00-12} \hbox{IHP-2000/05} \hbox{hep-th/0011215}} }
\pubtype{}
\date{November, 2000}

\titlepage

\title {\bf Symmetries and 
Compactifications of  (4,0) Conformal  Gravity}

\author{C.M. Hull}
\address{Physics Department,
Queen Mary and Westfield College,
\break
Mile End Road, London E1 4NS, U.K.}
\andaddress{Centre Emil Borel,
Institut Henri Poincar\' e,
\break
11 Rue Pierre et Marie Curie,
\break
75231 Paris Cedex 05, France.}
\vskip 0.5cm

\vskip 0.5cm

\abstract {The free (4,0) superconformal 
theory in 6 dimensions and its toroidal dimensional 
reductions are studied. The reduction to four dimensions on a 2-torus 
has an $SL(2,\Z)$ duality symmetry that acts non-trivially on the 
linearised gravity sector, interchanging the linearised Einstein 
equations and  Bianchi identities and giving a self-duality between 
strong and weak coupling regimes. The possibility of this extending 
to an interacting form of the theory  is discussed and 
implications for the non-geometric picture of gravity that could 
emerge are considered.}

\endpage

\chapter{Introduction}

In [\cst],  it was proposed that M-theory could have a superconformal phase
in $D=6$ dimensions with (4,0) supersymmetry  and $OSp^*(8/8) $ superconformal
invariance, arising from a strong coupling limit of five dimensional string
theory (i.e. M-theory compactified on a 6-torus).
The five-dimensional theory with $D=5$ Planck length $l$ is the
six-dimensional conformal theory compactified on a circle of radius $r=l$, so
that a limit in which $l$ becomes large is a decompactification limit in which
a new 6th dimension opens up. The $D=5$ Planck scale would thus have a geometric
origin. 

The free (4,0) theory
in six dimensions arises as
a  limit of the free $D=5$ linearised supergravity 
theory [\cst].
This, together with the analogous strong coupling limit of $D=5,N=4$ 
super-Yang-Mills to an interacting (2,0) superconformal field theory, 
motivated the conjecture of [\cst] that there could be a 
 strong coupling limit of the interacting $D=5$ 
supergravity theory to something that reduces to the free $D=6$ (4,0) 
theory in the linearised limit. A major obstacle in the checking of these 
conjectures is that the interacting (2,0) or (4,0) superconformal field 
theories -- if they exist -- cannot be formulated as local covariant 
field theories. One of the aims here is to seek indirect evidence for 
 these conjectures by investigating some of the consequences 
they would have, if true.

The (4,0) supermultiplet in $D=6$ has a 4th rank tensor gauge field $C_{MN\,
PQ}$ which has the same algebraic symmetry properties as the Riemann tensor,
and reducing to $D=5$ gives a graviton $h_{\mm\nn}$ only, with no other fields,
and the radius $r$ sets the $D=5$ Planck scale. The (4,0) multiplet reduces to
the $N=8$ supergravity multiplet in $D=5$. 
The $D=6$ theory has no graviton but has $C_{MN\, PQ}$ instead, and so 
the gravitational field
does not have an interpretation in terms of conventional Riemannian geometry.
A further reduction to $D=4$ on a circle of
radius $r'$ to an $N=8,D=4$ theory is 
straightforward; in particular, the $D=5$ graviton $h_{\mm\nn}$ gives a  graviton, a
vector and a scalar in $D=4$, with the scalar field $\phi$ given in terms of
the radius $r'$.
The two radii $r,r'$ appear to play rather different roles here, but the final
result might be expected to
be independent of which dimension is compactified first and so
should have a symmetry which interchanges the two radii.
The ratio $g^{2} = r'/r$ defines a dimensionless coupling constant (the
expectation value of $e^\phi$) and interchanging the two radii takes $g \to
1/g$, so that such a symmetry would be a duality interchanging weak and strong
coupling regimes, a kind of gravitational S-duality.

More generally, the reduction on a two-torus might be expected to give a $D=4$
theory with an $SL(2,\Z )$ duality symmetry
in addition to the $E_6(\Z)$ which is expected to be the duality symmetry of
the (4,0) theory.
However, the $D=4,N=8$ theory has an $E_7(\Z)$ U-duality symmetry 
[\HT] and 
$E_7$ does not have an $E_6\times SL(2)$ subgroup, so the $SL(2,\Z )$ duality
expected from the $D=6$ description could not be any of the known U-duality
symmetries of the theory [\jhs]. 
Moreover, such an $SL(2,\Z )$ duality would 
necessarily act non-trivially on the graviton, while the U-duality 
leaves the Einstein-frame metric invariant.
Such an $SL(2,\Z )$  symmetry would then necessarily include new 
symmetries not contained in the U-duality group.

One possibility is that there is indeed such a new $SL(2,\Z)$ symmetry in the
full theory, requiring some modification of the supergravity 
description.
This would then be similar to the case of the interacting (2,0) theory 
whose reduction to 4 dimensions has an $SL(2,\Z)$ S-duality symmetry. 
This is not a symmetry of the non-abelian $N=4$ super-Yang-Mills 
field equations, so that this field theory does not give a complete 
description of the S-dual theory.
Alternatively, the (4,0) theory is not based on conventional
geometry and so the geometric arguments leading to an $SL(2,\Z )$   symmetry
may be invalid. In particular, 
the $D=5$ diffeomorphisms do not arise from $D=6$ diffeomorphisms, but from a
higher-spin symmetry in $D=6$, and so the group $SL(2,\Z)$ of large
diffeomorphisms on a 2-torus need not play any role in the full theory.

In this paper, the  (4,0) theory in 6 
dimensions and its dimensional
reduction
 and  symmetries will be  investigated. Much of the paper will be devoted 
 to the free (4,0) theory, which is known explicitly, and
 its properties 
 have many similarities with the free (2,0) theory. The free (4,0) theory can be formulated in terms of free fields in a fixed flat
spacetime background. The bosonic fields consist of the gauge field $C_{MN\,
PQ}$,
27 2-form gauge fields $B_{MN}$ with self-dual field strengths and 
42 scalars.
If the background spacetime is the product of  4-dimensional Minkowski space
and a 2-torus, the free theory  has an $SL(2,\Z)$
symmetry arising from diffeomorphisms of the 2-torus background.
 For a 2-form gauge field reduced on a
2-torus, this is the S-duality of the resulting $D=4$ Maxwell theory, and this
is reviewed in section 2. In sections 3 and 4, the dimensional reduction of
the  gauge field $C_{MN\, PQ}$ and the resulting gravitational $SL(2,\Z)$
symmetry are studied.
As well as transforming photons into dual photons as in standard 
electromagnetic duality, it transforms gravitons into dual gravitons.
In section 5, 
it is shown how this occurs as part of the duality symmetry of the 
linearised $D=4$ supergravity.
In section 6, the question of
whether these symmetries can extend to an   interacting form of the theory 
(if it exists) is
addressed, leading to some insights into the possible structure of the 
theory.

\chapter{Dimensional reduction of 2-Form Gauge Fields}

The  abelian (2,0) theory  has a self-dual 
2-form
gauge field and five scalars, and its  dimensional reduction  
on a circle gives
$D=5$   super-Maxwell  theory (i.e. abelian super-Yang-Mills) and the reduction on a 2-torus gives 
the $D=4,N=4$   super-Maxwell theory 
with an $SL(2,\Z)$ S-duality symmetry arising from the 
mapping class group of the 2-torus [\Verl].
In this section, the dimensional reduction of both unconstrained and self-dual 
$D=6$ 2-form gauge  fields will be reviewed, with particular attention to  the
emergence of the duality symmetry.

A 2-form gauge field $B_{MN}$ in six dimensions
(where $M,N=0,1,...,5$) enjoys the gauge invariance
$$ \dd B_{MN}= \pa _{[M} \ll _{N]}
\eqn\abc$$
and the invariant field strength 3-form 
$H=dB$ satisfies the
 Bianchi identity
$$dH=0
\eqn\hbi$$
The conformally invariant six-dimensional free action
$$
S=-{1 \over 12}\int
d^6x \, \sqrt
{-g} H_{MNP}H^{MNP}
\eqn\hact$$
implies the field equation
$$ d*H=0
\eqn\hfi$$
(where $*$ denotes the Hodge dual,  $(*H)_{MNP}= \6 \epsilon _{MNPQRS}H^{QRS}
$ and $\epsilon _{MNPQRS}$ is a tensor, not a tensor density)
and the 2-form has 6 degrees of freedom in the $(3,1)+(1,3)$ representation  of the little
group
$SU(2)\times SU(2)$.

Imposing the
self-duality constraint
$$H=*H
\eqn\hsd$$
halves the number of degrees of freedom,  leaving 3 physical modes in
  the (3,1) of  $SU(2)\times SU(2)
$.
The self-duality condition \hsd\ together with  the Bianchi identity
 $dH=0$
  implies the field equation $d*H=0$.
The action \hact\ vanishes for self-dual $H$.

The dimensional reduction of a general 2-form gauge field
$B_{{MN}}$ from 6 to 5 dimensions on a circle of radius $r$  gives
a vector field $A_{\mm}=B_{\mm 5}$ with field strength $F=dA$
and a 2-form $B_{\mm\nn}$ with field strength $H=dB$ (where 
$\mm,\nn=0,1,\ldots4$). The reduction of the action \hact\  gives
$$
S=- \int
d^5x \, \left(  {1 \over 4g^2_{YM}}F_{\mu\nu}F^{\mu\nu}+ {g^2_{{YM}} \over  
6}H_{\mu\nu\rho}H^{\mu\nu\rho}
\right)
\eqn\abc$$
where the dimensionful  $D=5$ coupling constant $g_{{YM}}$ is given
by
$$g^2_{{YM}}=r
\eqn\abc$$
The 2-form $B_{\mm\nn}$ can be dualised to a second vector field $\ti A_\mm$
with field strength $\ti F=d\ti A$
given by
$$\ti F=r*H
\eqn\abc$$
so that the action becomes
$$S=-{1 \over 4g^2_{YM}} \int
d^5x \, \left(  F_{\mu\nu}F^{\mu\nu}
-
\ti F_{\mu\nu}\ti F^{\mu\nu}
\right)
\eqn\abc$$
Note that the Bianchi identity for $H$ \hbi\ implies the field equation for
$\ti A$ and vice versa.
 The 6 degrees of freedom of a general $B_{MN}$ in $D=6$ gives 
the 3 degrees of freedom of $A$ plus the 3 degrees of freedom of $B$ or $\ti A$
in $D=5$.

However, if the $D=6$ gauge field is self-dual, then \hsd\ implies the
$D=5 $ constraint
$$ F=r*H
\eqn\abc$$
so that 
$B$ is determined by $A$, or equivalently
$\ti F=F$ so that a gauge can be chosen in which
$\ti A=A$. Only 3 degrees of
freedom then  remain in $D=5$, which can be taken to be represented by the 
field $A$ with action 
$$ S=-  {1 \over 4g^2_{YM}}\int
d^5x \,  F_{\mu\nu}F^{\mu\nu}
\eqn\abc$$

The reduction of  $B_{MN}$ from 6  to 4 dimensions on a 2-torus with
constant metric $\gamma_{ij}$  gives two vector fields
$A_{mi}=B_{mi}$, a scalar field $\phi={1 \over 2}\epsilon^{ij} 
B_{ij}$ and a 2-form gauge field $ B_{mn}$,
where $m,n=0,3,4,5$ and $i,j=1,2$. (Here $\epsilon_{ij}$ 
has components $\epsilon_{12}= \sqrt{det \gg_{ij}}$ and the indices $i,j$
are raised and lowered with the metric  $\gamma_{ij}$.)
 The 2-form $B$ can be dualised to a
second scalar
$\ti
\phi$,  defined by
$$d\ti \phi = *H
\eqn\abc$$
with $H=dB$.
Then reducing the $D=6$ action \hact\ gives the 
$D=4$
 action  
$$
S= -\int
d^4x\, \left({1 \over 4}
\tilde \gamma^{ij} F_{mn\, i } F^{mn}{}_j +{1 \over 2} (\partial 
\phi)^2
+{1 \over 2} (\partial \tilde \phi)^2
\right)
\eqn\hactaa$$
where $$\tilde \gamma^{ij} =\sqrt
{\gamma} \gamma ^{ij}
\eqn\abc$$
This action is manifestly $SL(2,\R)$ invariant, with $F_{i}$ 
transforming as
$$F \to SF
\eqn\abc$$
where $S_i{}^j$ is an $SL(2,\R)$ matrix
and $\gg_{ij}$ transforms as 
$$ \gg \to S\gg S^{t}
\eqn\abc$$
The boundary conditions on the 2-torus coordinates or the
Dirac quantization conditions in $D=4$ break the 
$SL(2,\R)$ symmetry down to $SL(2,\Z)$.

Suppose now that
 $B_{{MN}}$ satisfies the self-duality constraint \hsd.
Dimensionally reducing \hsd\ gives
$$d  \phi = *H
\eqn\abc$$
and
$$ 
\tilde \gamma^{ij} F_{mn\, j}={1 \over 
2}\epsilon_{mnpq}\epsilon^{ij}F^{pq}{}_{j}
\eqn\fsd$$
so that $A_1,A_2$ are dual potentials corresponding to the same 
electromagnetic field, and $\phi $ is dual to $B$, so that
one can set $\ti \phi =\phi$.
The constraint \fsd\ can be written in the form used in [\CJP] as
$$F_i= J_i{}^j *F_j
\eqn\fsd$$
where
$$J_i{}^j
=\tilde \gamma_{ik}\epsilon^{kj}={1 \over \sqrt
{\gamma}}\gamma_{ik}\epsilon^{kj}
\eqn\abc$$
and satisfies $J^2=-1$.
The $3+3 $ degrees of freedom 
contained in $A_1,\phi,A_2,B$ then satisfy 3 constraints, leaving 
3 independent degrees of freedom which can be taken to be represented by
$A_1,\phi$.

\chapter{Tensor Gauge Fields}

\section{The Graviton}

It will be useful to review some of the properties of a free 
symmetric tensor gauge field $h_{MN}$ in $D$ dimensions before generalising to
the
fourth rank gauge fields of the type that occur in the (4,0)
multiplet.
It is useful to write the spacetime metric as
$$ g_{MN}=\eta _{MN}+ \ll h_{MN}
\eqn\abc$$
and the Einstein action as
$$S={1\over \ll^{2}l^{D-2}} \int d^{D}x \, \sqrt{-g} R
\eqn\abc$$
where $\ll$ is a dimensionless coupling constant.
If $\ll \ne 0$, then it can be absorbed into the definitions of 
$h_{MN}$ and $l$.
The two kinds of limit discussed in this paper are the
free limit in which $\ll \to 0$ leaving a free theory with a quadratic 
action for $h_{MN}$ and the strong coupling limit in which $l \to 
\infty$.

For the free theory, 
the gauge symmetry 
is
$$ \dd h_{MN}= \pa _{(M}\xi_{N)}
\eqn\abc$$
and the invariant field strength is the linearised Riemann tensor
$$ R_{MN\, PQ}=\pa_{M} \pa_{P}h_{NQ}+\ldots=-4\pa_{[M}h_{N][P,Q]}
\eqn\ris$$
This satisfies 
$$R_{MN\, PQ}=R_{PQ\,MN}
\eqn\abc$$
together with
the first Bianchi identity
$$R_{[MN\, P]Q} =0
\eqn\rbi$$
and the second Bianchi identity
$$\pa _{[S}R_{MN]\, PQ}=0
\eqn\rbii$$
The natural free field equation  in $D\ge 4$ is the linearised 
Einstein equation
$$R^P{}_{M\, PN}=0
\eqn\rfi$$
 which then implies 
$$\pa ^M R_{MN\, PQ}=0
\eqn\rfiij$$
where indices are raised and lowered with a flat background metric 
$\eta_{MN}$.
In $D=3$, the Weyl tensor vanishes identically so that the Riemann 
tensor is completely determined by the Ricci tensor, and  the field 
equation \rfi\ implies that 
the field strength \ris\ vanishes and $h_{{MN}}$ is pure gauge.
The simplest non-trivial linear field equation is
in $D=3$ is
$$R^{MN}{}_{MN}=0
\eqn\rfii$$
representing one degree of freedom. Indeed, in $D=3$, a free graviton 
satisfying the field equation \rfii\ can be dualised to a free scalar 
$\phi$ by the relation
$$R^{MN\, PQ}=\ee^{MNR}\ee^{PQS}\pa_{R}\pa_{S}\phi
\eqn\abc$$
with \rfii\ implying $\boxx \phi=0$.
In $D=2$, the  Riemann 
tensor is completely determined by the Ricci scalar, so that the field 
equation \rfii\ only has trivial solutions 
and there is no non-trivial linear field equation.

In four Euclidean dimensions, one can consistently impose the
self-duality condition
$$
R_{MN\, PQ}=\2 \ee_{MNST}R^{ST}{}_{PQ}
\eqn\abc$$
which implies the field equations \rfi, but is stronger.

\section{The 4th Rank Tensor Gauge Field }

   Consider  a
 gauge field $C_{MN\, PQ}
$  in $D$ dimensions with the algebraic properties of the Riemann tensor
$$ C_{MN\, PQ} =
-C_{NM\, PQ}
=
-C_{MN\, QP}
=
C_{PQ\, MN}, \qq C_{[MNP]Q}=0
\eqn\calg$$
and the gauge symmetry
$$ \dd C_{MN\, PQ} = \pa _{[M} \chi _{N]PQ} +\pa _{[P} \chi
_{Q]MN}-2
\pa _{[M} \chi _{NPQ]}
\eqn\delcis$$
with parameter $\chi _{MPQ}=-\chi _{MQP}$.
Such gauge fields were considered in four dimensions in [\notivarg] 
and arise in the (4,0) supermultiplet in $D=6$ [\cst].
The invariant field strength is
$$G_{MNP\, QRS} =\pa_{M}\pa_{Q}C_{NP\, RS}+\ldots =
9\pa _{[M} C_{NP]\,  [QR,S]}
\eqn\gis$$
so that
$$G_{MNP\, QRS} =G_{[MNP]\, [QRS]} =G_{QRS  \,   MNP}
\eqn\gisdfsds$$
This satisfies the first Bianchi identity
$$G_{[MNP\, Q]RS} =0
\eqn\gbi$$
and the second Bianchi identity
$$\pa _{[T}G_{MNP]\, QRS}=0
\eqn\gbii$$
The natural linear  free field equation in $D\ge 6$  is
$$G^Q{}_{NP\, QRS}=0
\eqn\gfi$$
which then implies (using \gisdfsds,\gbi)
$$\pa ^M G_{MNP\, QRS}=0
\eqn\gfii$$

In $D\le 5$, 
the trace-free part of the field strength vanishes identically, 
so that in $D=5$
$$G_{\mm\nn\rr}{}^{\ss\tt\ll}=9G_{[\mm\nn}{}^{[\ss\tt}\dd_{\rr]}{}^{\ll]}
-9
G_{[\mm}{}^{[\ss}\dd_{\nn\rr]}{}^{\tt\ll]}+G\dd_{\mm\nn\rr}{}^{\ss\tt\ll}
\eqn\abc$$
where
$$G_{\mm\nn}{}^{\ss\tt} = G_{\mm\nn\rr}{}^{\ss\tt\rr}, \qq
G_{\mm\ss}=G_{\mm\nn\ss}{}^{\nn},\qq G=G_{\mm}{}^{\mm}
\eqn\abc$$
(here 
$\mm,\nn=0,1,\ldots 4$).
This is analogous to the fact that
the trace-free part of the Riemann tensor, the Weyl tensor, vanishes in $D\le 3$.
The  field equation for $C_{\mm\nn\,\ss\tt }$ given by
$$
G_{\mm\nn\rr\,\ss\tt }{}^\rr=0
\eqn\fefga$$
then implies that the field strength vanishes,
 $G_{\mm\nn\rr\,\ss\tt\ll} =0
$, in dimensions $D\le 5$, so that 
$C_{\mm\nn\,\rr\ss}
$ is pure gauge.
Thus the field equation  \fefga\ with one contraction is trivial and the
simplest non-trivial linear 2nd order field equation is 
$$G^{\ss\tt}{}_{ \rr\,\ss\tt\ll}
=0\eqn\fefa$$
  with two contractions.
A gauge field $C_{\mm\nn\,\rr\ss}
$ satisfying this field equation in $D=5$ represents 5 degrees of 
freedom and can be  dualised to a linearised graviton [\cst]
$\hat h_{\mm\nn}$
  with
linearised curvature   $\hat
R_{\mm\nn\,
\rr\ss}(\hat h)$ given by
$$ 
\hat R_{\mm\nn\, \rr\ss}
={1\over 36} \ee _{\mm\nn \aa\bb\gg}G^{\aa\bb\gg\, \kk\ll\tt}{}
\ee _{  \rr\ss \kk\ll\tt}
\eqn\abc$$
or
$  \hat R=*G*$.
The Bianchi identities and field equation \fefa\ for $C_{\mm\nn\,\rr\ss}$
  then imply the Bianchi identities
  and linearised Einstein equation for $\hat R$.

A natural 2nd order field equation in $D=4$ 
for $C_{mnpq}$ ($m,n=0,1,2,3$) is
$$G^{mnp}{}_{mnp}=0
\eqn\geq$$
(The field equations given by requiring single or double contractions of $G$
to vanish are trivial in $D\le 4$ and force   $C$ to be pure gauge. The
simplest equation with non-trivial solutions is \geq\ with triple contractions
of $G$.)  If it satisfies this, it can be dualised to a scalar field $\hat
\phi$ defined by $$\pa_m\pa _n \hat \phi=(*G*)_{mn}
\eqn\gdu$$
In $D<4$, there is no non-trivial linear field equation for such 
gauge fields.

In [\notivarg], a $D=4$ gauge field $C_{mnpq}$ with the same algebraic
properties \calg\ was
  considered, and there it was shown that an unusual higher derivative field
equation for   $C_{mnpq}$ gave a system that was dual to a Maxwell  vector field.
The name notivarg was proposed for $C_{mnpq}$  in [\notivarg], in analogy with
the name notoph proposed in [\notoph] for a 2-form gauge field $B_{mn}$ 
in four dimensions.
Here, the 2nd order field equation \gfi\ implies that $C_{mnpq}$ is dual to a scalar,
not a vector field.

The gauge field $C_{{MN\,PQ}}$ in $D=6$ with field equation \gfi\   represents
10 degrees of freedom in the $(5,1)+(1,5)$ representation  of the little
group
$SU(2)\times SU(2)$ [\cst].
In $5+1$ dimensions, one can consistently impose the 
the self-duality constraint
 $$G=*G
\eqn\gsd$$
which then implies $G=G* =*G*$
where
$$\eqalign{(*G)_{MNP\, QRS}
&= \6 \epsilon _{MNPTUV}
G^{TUV}{}_{ QRS}
\cr
 ( G*)_{MNP\, QRS}
&= \6 \epsilon _{QRS   TUV}
G_{MNP} {}^{TUV}{} \cr}
\eqn\abc$$
This halves the degrees of freedom to 5 in the $(5,1) $ representation  of the
little group.
The 4-th rank gauge field in the
  the (4,0) multiplet 
  satisfies such a self-duality constraint [\cst].
The self-duality constraint \gsd\ and the Bianchi identities \gbi,\gbii\   
imply the field equations \gfi,\gfii.

\section {Reduction from Six to Five Dimensions}

Reducing a general unconstrained 4th rank  $D=6$ gauge field  $C_{MN\, PQ}$ to 
$D=5$ on a circle of radius $r$
gives
the fields
$$h_{\mm\nn}=C_{\mm 5\, \nn 5},
\qq
D_{\mm\nn\,\rr}=C_{\mm\nn\,\rr 5},\qq
 C_{\mm\nn\,\rr\ss}
\eqn\abc$$
with the algebraic properties
 $$h_{[\mm\nn]}=0,
\qq
D_{[\mm\nn\,\rr]}=0,\qq
 C_{[\mm\nn\,\rr]\ss}=0
\eqn\abc$$
The linear field strengths $R_{\mm\nn\, \rr\ss}$ for $h_{\mm\nn}$,
$S_{\mm\nn\rr\, \ss\tt}$ for $D_{\mm\nn\rr}$ and 
$G_{\mm\nn\rr\,\ss\tt\ll}$ for $C_{\mm\nn\,\rr\ss}$
are given by
$$\eqalign{&R_{\mm\nn\, \rr\ss}=G_{\mm\nn 5\, \rr\ss 5}
\cr &
S_{\mm\nn\rr\, \ss\tt}=G_{\mm\nn\rr\, \ss\tt 5}
\cr &
G_{\mm\nn\rr\,\ss\tt\ll} 
\cr}
\eqn\abc$$
so that
$$
 S_{\mm\nn\rr\, \ss\tt} =-6 \pa _{[\mm} D_{\nn\rr]\,  [\ss,\tt]},\qq
 G_{\mm\nn\rr\,\ss\tt\ll}=9\pa_{[\mm}C_{\nn\rr]\,[\ss\tt,\ll]}
\eqn\abc$$
and 
$R_{\mm\nn\, \rr\ss}$ is the linearised Riemann tensor for 
$h_{\mm\nn}$, given by \ris.

Consider first
the natural free field equations in $D=5$ for the gauge fields
$h_{\mm\nn},
D_{\mm\nn\,\rr},
 C_{\mm\nn\,\rr\ss}$
that are given by
$$\eqalign{&R^\rr{}_{ \nn\, \rr\ss}=0
\cr &
S^\ss{}_{ \nn\rr\, \ss\tt}=0
\cr &
G^{\ss\tt}{}_{ \rr\,\ss\tt\ll}
=0
\cr}
\eqn\fef$$
(As discussed above, the  field equation for $C$ given by
$
G_{\mm\nn\rr\,\ss\tt }{}^\rr=0$
that might have been expected by comparison to the $D=6$ equation 
\gfi\
would be trivial, so \fefa\ is the simplest non-trivial possibility.)
If the field equations \fef\ are satisfied, then  each of the gauge fields
represents 5 degrees of freedom. Indeed, it was seen in [\cst] that in the
linearised $D=5$ theory a linearised graviton
$h_{\mm\nn}$ has dual
representations in terms of
fields $D_{\mm\nn\,\rr}$ or $C_{\mm\nn\,\rr\ss}
$, so 
$C_{\mm\nn\,\rr\ss}$ can be dualised to give a 2nd 
graviton  (i.e. free symmetric tensor gauge field) $\hat
h_{\mu\nu}$ and $
D_{\mm\nn\,\rr}$ can be dualised to a third one $\tilde h_{\mu\nu}$.
Linearised gravitons $\ti h_{\mm\nn}$ and $\hat h_{\mm\nn}$
which are dual to $D_{\mm\nn\rr}$ and   $C_{\mm\nn\,\rr\ss}$
respectively are introduced by requiring that the corresponding
linearised curvatures $\tilde R_{\mm\nn\, \rr\ss}(\ti h)$ and $\hat
R_{\mm\nn\,
\rr\ss}(\hat h)$ satisfy
$$\eqalign{&\tilde R_{\mm\nn\, \rr\ss}
=r\6 \ee _{\mm\nn \aa\bb\gg}S^{\aa\bb\gg}{}_{  \rr\ss}
\cr &
\hat R_{\mm\nn\, \rr\ss}
=r^{2} {1\over 36} \ee _{\mm\nn \aa\bb\gg}G^{\aa\bb\gg\, \kk\ll\tt}{}
\ee _{  \rr\ss \kk\ll\tt}
\cr}
\eqn\abc$$
or
$$ \ti R = r*S,\qq \hat R=r^{2} *G*
\eqn\abc$$
provided the field equations are those in \fef.
Then the Bianchi identities and  field equations  of each field imply 
those
of its dual.

If the reduction to $D=5$ of a $D=6$ gauge field satisfying \gfi\
gave the three gauge fields $h,D,C$ with the field
equations \fef, there would be 15 degrees of freedom and the resulting system
could be dualised to one with three free gravitons
$h,\ti h, \hat h$. However, there were only 10 degrees of freedom in $D=6$, so 
the dynamics here must be different.
In fact, the reduction   of the $D=6$ gauge field
gives stronger field equations than \fef\  and only two independent gravitons
remain, as will now be shown.

The reduction of the $D=6$ field equation \gfi\ gives the first two field
equations of \fef, but the field equation for $C$ is
$$
G_{\mm\nn\rr\,\ss\tt }{}^\rr= -{1\over r^{2}}
R_{\mm\nn \,\ss\tt }
\eqn\fefg$$
This implies the \lq natural' field equation $G^{\ss\tt}{}_{
\rr\,\ss\tt\ll} =0$ of \fef\ (using the first field equation $R_{\mm\nn}=0$),
but is stronger. 
Since \fefg\ fixes the trace of $G_{\mm\nn\rr\,\ss\tt\ll} 
$, it completely determines the whole of $G_{\mm\nn\rr\,\ss\tt\ll} 
$ in terms of $R_{\mm\nn \,\ss\tt }$:
$$G_{\mm\nn\rr}{}^{\aa\bb\gg}=-9 {1\over r^{2}}
R_{[\mm\nn}{}^{[\aa\bb}\dd_{\rr]}{}^{\gg]}
\eqn\abc$$
(using $R_{\mm\nn}=0$).
This implies that $C$ is not an independent degree of freedom, and can be solved
for in terms of $h$ as 
$$C_{\mm\nn}{}^{\rr\ss}=-4{1\over r^{2}}h_{[\mm}{}^{[\rr}\dd_{\nn]}{}^{\ss]}
\eqn\abc$$
up to gauge transformations, 
or equivalently after the field redefinition
$$C_{\mm\nn}{}^{\rr\ss}\to \bar
C_{\mm\nn}{}^{\rr\ss}=C_{\mm\nn}{}^{\rr\ss}+
4{1\over r^{2}} h_{[\mm}{}^{[\rr}\dd_{\nn]}{}^{\ss]}
\eqn\abc$$
the field $\bar C$ has the field equation
$$
\bar G_{\mm\nn\rr\,\ss\tt }{}^\rr= 0
\eqn\abc$$
implying that $\bar C$ is pure gauge and so trivial.
In terms of the dual variables, this implies that the graviton 
$\hat h_{\mm\nn}$ (dual to $C$) satisfies
$\hat R_{\mm\nn\ss\rr} =R_{\mm\nn\ss\rr}$ so that 
(up to gauge transformations)
$  \hat h_{\mm\nn}=h_{\mm\nn}$. Then  the system has only 10 independent degrees of
freedom, represented by the 5 of the graviton 
$  h_{\mm\nn}$ and the 5 of  $\ti  h_{\mm\nn}$ or its dual $D_{\mm\nn\,\rr}$.

Suppose now that the $D=6$ gauge field is self-dual, satisfying  \gsd.
This gives the $D=5$ constraints
$$  R =r *S=r^{2} *G*
\eqn\abc$$
which 
 implies that $C,D$ are not independent degrees of freedom
but can be eliminated in terms of $h$.
 In terms of the dual variables, this implies that (up to gauge 
 transformations)
$$\hat h_{\mm\nn}= \ti h_{\mm\nn}= h_{\mm\nn}
\eqn\abc$$
so that there is only one independent   graviton and only 5 degrees of freedom,
as required.

\section {Reduction from Six to Four Dimensions}

Reducing the gauge field $C_{MNPQ}$ to $D=4$ on a 2-torus with 
constant metric
$\gg_{ij}$ gives the gauge fields and their corresponding field strengths
$$\eqalign{&C_{mnpq} ,\qq \qq G _{mnpqrs} = \pa_{m} \pa _{q} C_{nprs}+\ldots
\cr &
D_{mnp\, i}=C_{mnpi},\qq S_{mnpqr \, i} = \pa_{m} \pa _{q} D_{npr\, i}+\ldots
\cr &
B_{mn}=\2 \ee ^{ij}C_{mn ij},\qq H=dB
\cr &
h_{mn\, ij}= -C_{m(i j) n},\qq R_{mnpq\, ij}=\pa_{m} \pa _{p} h_{nq\, 
ij}+\ldots
\cr &
A_{mi}=\2 \ee ^{jk}C_{m  ijk},\qq F_i=dA_i
\cr &
\phi=\4 \ee ^{ij} \ee ^{kl}C_{   ij \, kl},\qq P=d\phi
\cr}
\eqn\abc$$
where, as in section 2,
  $\epsilon_{ij}$ 
has components $\epsilon_{12}= \sqrt{det \gg_{ij}}$ and the indices $i,j$
are raised and lowered with the metric  $\gamma_{ij}$.
Dimensionally reducing the field strength
$G_{MNPQRS}$
gives
$$\eqalign{&  G _{mnp\, qrs} 
\cr &
G_{mnp\, qr   i} = S_{mnpqr \, i}
\cr &
G_{mnp\, q    ij} = \pa_q H_{mnp}\ee_{ij}
\cr &
G_{mn    (ij)pq} = R_{mnpq\, ij}
\cr &
G_{mni \, p      jk} =   \pa_p F_{mn i}\ee_{ jk}
\cr &
G_{mij\, nkl}  =\ee_{ij}\ee_{kl}
\pa_m\pa _n \phi
\cr}
\eqn\abc$$

The usual free field equations for the 0,1,2 form gauge fields $\phi,A_i,B$
are
$$d*P=0, \qq d*F=0, \qq d*H=0
\eqn\feq$$
and the 2-form $B$ can be dualised to a scalar
$\ti \phi$
defined by
$$ d\ti \phi =*H
\eqn\hduu$$
The natural free 2nd order field equations for the three symmetric tensor
gauge fields
$h_{mn\, ij}= h_{(mn)\, (ij)}$ are the linearised Einstein equations
$$R^p{}_{ npq\, ij}
=0\eqn\ein$$

As above, the natural 2nd order field equation in $D=4$ 
for $C_{mnpq}$ is
$$G^{mnp}{}_{mnp}=0
\eqn\geq$$ which implies it can be dualised to a scalar field $\hat
\phi$ defined by $$\pa_m\pa _n \hat \phi=(*G*)_{mn}
\eqn\gdu$$
Similarly, the natural 2nd order field equation in $D=4$ 
for $D_{mnp\, i}$ is
$$S^{mn }{}_{p\, mn \, i}=0
\eqn\seq$$
and if it satisfies this, $D_{mnp\, i}$ can be dualised to vector fields
$\ti A_{mi}$
through the relation
$$ S_{mnp\, qr\, i}= \ee_{mnpt} \pa ^t \ti F_{qr\, i}
\eqn\abc$$

Reducing the $D=6$ field equation \gfi\ gives the equations
\feq\ but the equations for the remaining gauge fields are different from
\ein,\geq,\seq. The field equation  
for $D_{mnp\, i}$ obtained by reduction is
$$S^{m  }{}_{np\, mq \, i}= -\ee _{ij}
\pa _q F_{np}{}^j
\eqn\seqq$$
This implies \seq, but is stronger.
The trace-free part of $S_{mnpqri}$ vanishes identically in $D=4$, so that
\seqq\ determines   the field strength $S_{mnpqri}$ completely in terms of
$A_i$, and  implies that
$$D_{mn\, pi} \propto \ee_{ij}\eta _{p[m}A_{n]}{}^{j}
\eqn\dis$$
up to gauge transformations, so that $D$ does not represent independent degrees
of freedom.
In terms of the dual potentials $\ti A_i$, 
the constraint implies
$$ \ti F_i= \ee _{ij} (*F)^j
\eqn\ftis$$
so that the potentials $\ti A_i$ are dual to $\ee_{ij}A^{j}$.

The remaining field equations are
$$R^p{}_{ npq\, ij}
=-\gg_{ij}\pa_n\pa_q \phi
\eqn\hreq$$
and
$$G^r{}_{mn\, rpq}=-\gg^{ij} R_{mnpq\, ij}
\eqn\greq$$
and   these imply \ein,\geq, but are stronger.
It will be useful to decompose the 
three gravitons
as 
$$h_{mn\, ij}= g_{mn}\gg_{ij}+ h'{}_{mn\, ij}
\eqn\abc$$
where
$\gg^{ij}h'{}_{mnij}=0$.
As in $D=5$, \greq\ determines $C_{mnpq} $ completely in terms of the
graviton
$g_{mn}\equiv \2 \gg^{ij}h_{mnij}$.
Furthermore, the equations \hreq,\greq\ determine the curvature
of $g_{mn}$, defined by
$$R{}_{mnpq}=\2 \gg^{ij}R_{mnpqij}\eqn\abc$$
completely in terms of $\phi$.
Then $C$ and $g_{mn}$
are constrained to be
$$\eqalign{C_{mnpq}&={2\over 3} \eta_{m[p}\eta_{q]n} \phi
\cr
g_{mn}&=-\2 \eta_{mn}\phi
\cr}
\eqn\ciss$$
up to gauge transformations.
In terms of the scalar field $\hat \phi$ dual to
$C_{mnpq}$, the field equations imply 
$$\hat \phi = \phi
\eqn\phiss$$

To summarise, on dimensionally reducing the 
gauge field $C_{MNPQ}$ satisfying \gfi,
there are
 three scalars $\phi, \ti \phi,\hat \phi$ satisfying one constraint \phiss\
leaving two
independent scalars, 
there are four vector fields $A_i, \ti A_i$ satisfying two constraints 
\ftis\
leaving two independent vector fields and there are 
three symmetric tensor 
gauge fields $h{}_{mn\, ij}$ satisfying one constraint given by \greq\ or 
\ciss,
leaving two
independent ones.
 Then the  independent degrees of freedom can be taken to be  represented by
the two scalars $\phi,\ti \phi$, the two vector fields $A_i$ and the two
gravitons $h'{}_{mn\, ij}$, which are a symmetric traceless tensor in the
$i,j$ indices, giving a total of 10 degrees of freedom, as required.
The field equations are manifestly invariant under $SL(2,\R)$, with the two
vector fields
$A_i$ transforming as a doublet and the gravitons $h _{mn\, ij}$
transforming as a symmetric tensor, subject to the $SL(2,\R)$-invariant
constraint that they be trace-free, $\gg^{ij}h _{mn\, ij}=0$.

The $D=6$  self-duality condition \gsd\ halves the number of degrees of
freedom, leaving in $D=4$ one 
 scalar, one vector field and one
graviton.
The constraint \gsd\ implies, on reduction, that
$$\eqalign{&  G _{mnpqrs} = \ee_{mnpt}\ee_{pqru}\pa^t\pa^u\phi
\cr &
 S_{mnpqr \, i}= \ee_{mnps}\pa^sF_{qr\, i}
\cr &
 \pa_q H_{mnp}=\ee_{mnps}\pa^s \pa_q\phi
\cr &
R_{mnpq\, ij}= (*R)_{mnpq\, k(j}J_{i)}{}^k-\gg_{ij}\ee_{mnrs}\pa^r H^s{}_{pq}
\cr &
  \pa_p F_{mn i}= J_i{}^j\pa_p(* F)_{mn j}
\cr}
\eqn\abc$$
As before, these determine $C_{mnpq}$ or $\hat \phi$  in terms of
$\phi$ \ciss,\phiss\ and $D_{mnp\, i}$ or $\ti A_i$ 
in terms of 
$A_i$ \dis,\ftis.
In addition, they imply
$$ H=*d\phi
\eqn\hisphi$$
(up to the addition of a   constant 3-form which could arise as a constant
of integration) so that $B$ is dual to $\phi$, or equivalently
$$ \phi = \ti \phi
\eqn\abc$$
so that now $\phi = \ti \phi=\hat \phi$ and 
only one of the three scalars is independent.
Furthermore, the two 1-form  gauge fields satisfy the duality constraint
$$   F_{mn i}= J_i{}^j (* F)_{mn j}
\eqn\abc$$
(again suppressing a possible constant 2-form)
so that $A_2$ is dual to $A_1$, and only one  of the   1-form  gauge
fields is independent.

The constraint on the curvature tensors again determines 
$g_{mn}$ in terms of $\phi$ through \ciss\ (using \hisphi).
Equivalently, the field redefinition
$$h_{mn\, ij}\to \bar  h_{mn\, ij}= h_{mn\, ij} + \gg_{ij} \eta_{mn}\phi
\eqn\rede$$
brings the 
constraint to the form
$$
\bar R_{mnpq\, ij}= J_{i}{}^k(*\bar R)_{mnpq\, kj}
\eqn\rcon$$
or 
$$ \bar R=  J*\bar R  
\eqn\abc$$
This  implies that
$$
\gg^{ij}\bar R_{mnpq\, ij}= 0
\eqn\rtr$$
so that (after the redefinition
\rede)  $\bar g_{mn}=\2 \gg^{ij}\bar h_{mn\, ij}$ is trivial and can 
be gauged to zero.
This also implies that the right hand side of \rcon\ is symmetric in the
indices $i,j$
and that 
$$
\bar R_{mnpq\, ij}= J_{i}{}^k(\bar R *)_{mnpq\, kj}
\eqn\rcona$$
The 1st Bianchi identities
$$\bar R_{[mnp]q\, ij}=0
\eqn\abc$$
and the self-duality condition \rcon\ then imply the
field equations
$$\eta^{mp}\bar R_{mnpq\, ij}=0
\eqn\abc$$
Moreover, \rcon\ gives a duality relation between the two
remaining  gravitons, so that only one independent graviton remains.
Then the $D=6$  self-duality condition \gsd\ indeed halves the number of
degrees of freedom, leaving
  one 
 scalar, one vector field and one
graviton in $D=4$ (which is of course what is to be expected from the
reduction of the $D=5$ graviton $h_{\mm\nn}$).
However,
the self-duality conditions \rcon,\rtr\ are $SL(2,\R)$ invariant, so that the
classical $D=4$
theory has an $SL(2,\R)$ symmetry.

\chapter{Gravitational S-Duality in Four Dimensions}

In this section, the $SL(2,\Z)$ duality will be discussed both for 
the general systems obtained by reducing \hact\ or \gfi, and for the 
self-dual systems satisfying \fsd\ or \rcon.
The 2-torus metric can be written in terms of a complex modulus
$$ \tau =\tt_1+i\tt_2= \th + {i\over g^2}
\eqn\abc$$
  and the volume $V$ as
$$ \gamma_{ij}={V\over \tt_2}\pmatrix{
1 & \tau_1 \cr
\tt_1 & |\tt|^2 \cr
}
\eqn\abc$$
The $SL(2)$ invariant action \hactaa\ for the doublet  $(F_1,F_2)$ of 
independent field strengths is
$$S=\int d^{4 }x {1 \over \tt_{2}}
\left(
\vert \tt \vert ^{2 }(F_{1})^{2}+(F_{2})^{2}-2 \tt_{1 }F_{1}\cdot F_{2}
\right)
\eqn\fgfdgfdgagejfjk$$
and reduces to
$$S=\int d^{4 }x 
\left({1 \over g^{2}}
 (F_{1})^{2}+g^{2}(F_{2})^{2} 
\right)
\eqn\jkhdkjhd$$
for rectangular tori with  $\th =0$.
The self-duality equation for $F_i$ \fsd\ (which implies the 
vanishing of the action \fgfdgfdgagejfjk) gives
$$F_2= {1\over g^2}*F_1 + \th F_1
 \eqn\abc$$

It will be convenient to parameterise the triplets of metric and curvature 
tensors (after the redefinition \rede)  as
$$
\bar h_{mn\, ij}= \pmatrix{
h_{mn}  & \ti h_{mn}  \cr
\ti h_{mn}& \hat h_{mn} \cr
}, \qq \bar
R_{mnpq\, ij}= \pmatrix{
  R_{mnpq } & \ti R_{mnpq }  \cr
\ti R_{mnpq }& \hat R_{mnpq } \cr
}
\eqn\abc$$
Then \rtr\ implies
$$ \hat R_{mnpq } -2 \tt_1 \ti R_{mnpq }+ |\tt|^2 R_{mnpq }=0
\eqn\rrtr$$
so that
$$ \hat h _{mn} -2 \tt_1 \ti h _{mn}+ |\tt|^2 h _{mn}=0
\eqn\abc$$
up to gauge transformations.

It is straightforward to find the quadratic action implying the field 
equation \gfi.
In physical gauge, with transverse traceless $C_{MNPQ}$ [\cst]
the action simplifies to
$$\int d^{6}x\sqrt {-g}\, \, C_{MNPQ}\boxx  C^{MNPQ} 
\eqn\abc$$
Reducing on a 2-torus, this gives
the following physical gauge action for the gravitons $\bar h_{mn\, ij}$, which are 
also transverse and traceless:
$$S={1\over 4V}\int d^{4}x \, \ti \gg ^{ik}\ti \gg^{jl}
\bar h_{mnij}\boxx \bar h^{mn}{}_{kl}
\eqn\abc$$
which is manifestly $SL(2)$ invariant. If $\th =0$ this action is
$$S={1\over 4V}\int d^{4}x \left({1 \over g^{4}}
h_{mn}\boxx h^{mn}
+
2\ti h_{mn}\boxx\ti  h^{mn}
+g^{4}
\hat h_{mn}\boxx \hat h^{mn}
\right)
\eqn\kfkdfa$$
while the constraint \rtr\ becomes 
$$\hat h =-{1 \over g^{4}}h
\eqn\abc$$
and imposing this reduces the action \kfkdfa\ for the two gravitons $h,\ti 
h$ to
$$S={1\over 2V}\int d^{4}x \left({1 \over g^{4}}
h_{mn}\boxx h^{mn}
+
\ti h_{mn}\boxx \ti  h^{mn}
\right)
\eqn\abc$$
The covariant form of the action \kfkdfa\ is the sum of the linearised 
Einstein-Hilbert actions for $h,\ti h, \hat h$.

 The duality constraint \rcon\
gives the following relations between the curvature tensors (suppressing
the indices)
$$\eqalign 
{R&={1\over \tt_2}( -*\ti R +\tt_1*  R)
\cr
\ti R&={1\over \tt_2}( -* \hat R +\tt_1 * \ti  R)
\cr
\hat
R&={1\over \tt_2}( |\tt|^2*\ti R -\tt_1  *\hat R)
\cr}
\eqn\rrcon$$
This implies that $\ti R,\hat R$ are given in terms of $R$ by
$$\eqalign{\ti R&= {1\over g^2}*R - \th R
\cr 
\hat R&= 2 \tt_{1}\tt_{2}*R - (\tt_{1}^{2}-\tt _{2}^{2}) R
\cr}
 \eqn\abc$$
 The one remaining independent graviton can be taken to be $h$ with 
 $\th=0$
 action
 $$S={1\over 2l^{2}}\int d^{4}x   \, 
h_{mn}\boxx  h^{mn}
\eqn\abc$$
where 
the Planck length is given by
$$l=\sqrt{V} g^{2}
\eqn\abc$$
The reduction gives two dimensionless 
coupling constants, $g,\th$. While $g$ can be  absorbed into the 
the gravitational coupling $l$, there is the interesting possibility 
of a gravitational $\th$-parameter.
 
Under the action of $SL(2)$, 
$\bar R_{mnpq\, ij}$ transforms as a symmetric tensor 
$$\bar R_{mnpq } \to S\bar R_{mnpq }S^{t}\eqn\symten$$
with the invariant tracefree
condition \rtr, $F_i$ transforms as a vector $F\to SF$ and $\tt$ 
transforms through
a fractional linear transformation.
For $F_{i}$ satisfying the self-dual condition \fsd,
 an $SL(2)$ transformation takes $F_{1}$ to a 
 linear combination of $F_{1}$ and $*F_{1}$ while for 
 curvatures satisfying the self-duality condition \rcon\ 
 it takes $R$ to  a 
 linear combination of $R$ and $*R$
 and   mixes the field equations
$$\eta ^{mp}R_{mnpq}=0
\eqn\abc$$
with the Bianchi identities
$$\eta ^{mp}(*R)_{mnpq}=0
\eqn\abc$$

The action of the $SL(2,\Z)$ element
$$S=\pmatrix{
0 & 1 \cr
-1 & 0\cr
}\eqn\abc$$
is to take 
$$\tt \to - {1\over \tt}
\eqn\abc$$
and 
$$ F_1 \to F_2, \qq F_2 \to - F_1
\eqn\abc$$
while 
$$ R \to \hat R, \qq \hat R \to R , \qq \ti R \to - \ti R
\eqn\abc$$
Note that this preserves the constraint \rrtr.
For self-dual $F_{i}$, the transformation is the standard duality 
transformation
$$F_{1}\to {1\over g^2}*F_1 + \th F_1
\eqn\abc$$
while for self-dual $\bar R$
$$R_{mnpq} \to
2 \tt_{1}\tt_{2}(*R)_{mnpq} - (\tt_{1}^{2}-\tt _{2}^{2}) R_{mnpq}
\eqn\abc$$
For $\th=0$, this $SL(2,\Z)$ transformation 
takes $g \to 1/g$ and so relates strong and weak 
coupling regimes. Both the linear gravity and Maxwell theory are 
self-dual, with the strong coupling regime described by the same 
theory.

For Maxwell theory, maintaining duality in the presence of sources 
requires introducing a magnetic current $\ti J$ as well as an electric current
$J$
with
$$ dF=*\ti J,\qq d*F= *J
\eqn\abc$$
and $(J,\ti J)$ forming an $SL(2)$ doublet.
In regions in which $\ti J$ vanishes, $F$ can be solved for in terms 
of a potential $A$, $F=dA$, while
in regions in which $ J$ vanishes, $*F$ can be solved for in terms 
of a dual potential $\ti A$, $*F=d\ti A$.
Similarly, duality requires introducing 
energy-momentum tensors $T_{mn ij}$ for all three gravitons $h_{mn ij}$
in the general system \kfkdfa, while for the self-dual system satisfying 
\rcon, this requires introducing
a \lq magnetic' energy-momentum tensor $\ti T_{mn }$
as well as the usual $ T_{mn }$
with
$$\eta ^{mp}R_{mnpq}= T_{nq}+\2 \eta_{nq} T
\eqn\abc$$
and a source for the first  Bianchi identity
$$\eta ^{mp}(*R)_{mnpq}=\ti  T_{nq}+\2 \eta_{nq} \ti T
\eqn\abc$$
In regions in which $\ti T_{mn }$ vanishes, $R_{mnpq}$ can be solved for in terms 
of a graviton $h_{mn}$, while
in regions in which  $ T_{mn }$ vanishes, $\ti 
R_{mnpq}$  can be solved for in terms 
of a dual  graviton $\ti h_{mn}$.

\chapter{Symmetries of the Linear (4,0) Theory and its Compactifications}

The maximal supergravity theory in $D$ dimensions has a rigid duality
symmetry $G$ which is broken to a discrete U-duality subgroup $G(\Z)$ in
the quantum theory; for $D=4$, $G=E_{7(+7)}$ [\CJ] and for $D=5$, 
$G=E_{6(+6)}$ [\Cremmer]. The scalar
fields take values in the coset space $G/H$ where $H$ is the maximal
compact subgroup of $G$, and the theory can be formulated with local $H$
symmetry [\CJ,\julia]. However, the free limit of these supergravities in general have 
much larger symmetry groups and the symmetry groups of the linearised theories
will play a role here.

The scalar fields in a supergravity theory take values in some target space
{\cal M}.
The non-linear sigma-model action   has an invariance under the diffeomorphism
group of {\cal M}, the group
$Diff ({\cal M})$ of pseudo-symmetries or sigma-model symmetries.
The isometry subgroup of this $Isom  ({\cal M})$ 
are proper symmetries of the action. 
In maximal supergravity theories, {\cal M} is 
a coset space $G/H$ and the isometry group is $G$, giving the rigid $G$ duality
symmetry. For $n$ free scalar fields, the target space is $n$-dimensional flat
space and the isometry group is
the Euclidean group  $IO(n)=O(n) \xx \R^n$ of translations, reflections and rotations.
The free scalar field equations
have the larger symmetry $IGL(n)=GL(n) \xx \R^n$.
The $N=8,D=5$ supergravity has 42 scalar fields
in $G/H=E_6/USp(8)$ and in the free limit the symmetry $E_6$ of the 
scalar kinetic term is increased to
$IGL(42)$, which contains the group contraction of $E_6$ given by
$USp(8)\xx \R^{42}$ as a subgroup. In the full theory,
the scalar self-interactions 
break the symmetries not in this subgroup and
modify the translation symmetries so that
$USp(8)\xx \R^{42}$ becomes $E_6$.
The situation is similar for $N=8,D=4$ supergravity which  has 70 scalar fields
in $G/H=E_7/SU(8)$ and in the free limit the symmetry $E_7$ is increased to
$IGL(70)$, which contains the group contraction of $E_7$ given by
$SU(8)\xx \C^{35}$ as a subgroup. 

The $D=5,N=8$ supergravity has 27 vector gauge fields and so the classical
free vector field equations have a $GL(27,\R)$ symmetry, broken to $GL(27,\Z)$
in the
quantum  theory. The
 interactions with the scalars break the $GL(27)$ symmetry to $E_6$
with the vector fields transforming as a {\bf 27}.
The $D=4,
N=8$ supergravity has 28
vector gauge fields with field strengths
$F_A$, $A=1,...,28$, satisfying $dF_A=0$. The vector
field equations can be written as 
$dG_A
=0$ where the 28 dual field strengths $G_A$ are given by
$G_A=*F_A$ in the free theory, while in the interacting theory 
$G_{A}=*(\dd S/\dd F_{A})= *F_A+\ldots$. 
The field equations imply that the $G_A$ can be written in terms of dual
potentials $\ti A_A$, $G_A=d\ti A_A$.
The 56-vector of field strengths given by
$${\cal F} = 
\pmatrix{
\hat F_A
 \cr
\hat G_A\cr
}
\eqn\abc$$
then satisfies (setting the fermions to zero) [\CJ]
$${\cal F} = J*{\cal F} , \qq d {\cal F}=0
\eqn\cjeq$$
where $J$ is the $56 \times 56 $ matrix
$$J ={\cal V}^{-1} \www {\cal V} ,\qq \www\equiv \pmatrix{
0 & 1 \cr
-1 & 0\cr
}\eqn\jiss$$
and ${\cal V} (\phi)$ is the scalar-dependent 56-bein parameterising  the 
scalar coset space.
In the free case, $J=\www$ and the equations \cjeq\ have
an $Sp(56,\R)$ symmetry (the $56
\times 56 $ matrices preserving $\www$) [\GZ]
which is broken to $Sp(56,\Z)$ in the quantum theory. In the full 
theory the interactions with
the scalar fields break the $Sp(56)$ to $E_7$ and these equations are 
  supercovariantized with   fermion bilinears [\CJ].

Finally, the linearised Einstein equations in $D=4$ can be expressed in terms
of the matrix of dual gravitons  as \rcon\ with a manifest $SL(2) $ symmetry, although
this is not a symmetry of the non-linear Einstein equations.
Thus the classical free bosonic  field equations
of linearised $N=8,D=4$ supergravity have a symmetry
$$SL(2) \times Sp(56) \times IGL(70)
\eqn\fsym$$
while those of linearised
$N=8,D=5$ supergravity have a symmetry
$$  GL(27)\times IGL(42)
\eqn\gsym$$

The bosonic field equations of the free (4,0) theory
$$ G=*G, \qq H_{a}=*H_{a}, \qq \boxx \phi^{i}=0
\eqn\abc$$
also have the symmetry \gsym, with
the 42 scalars $\phi^{i}$ taking values in $\R ^{42}$ and acted on by   
    $IGL(42)$ while the 27 self-dual 3-form   field 
strengths $H_{a}$
transform as
a {\bf 27} of $GL(27)$.

The dimensional reduction of the free (4,0) theory on a 2-torus 
must have the symmetry \gsym\ together with an extra $SL(2)$ symmetry 
from the 
reparameterizations of the 2-torus, and
$$ SL(2) \times GL(27)\times IGL(42)
\eqn\hsym$$
 is indeed 
a subgroup of the symmetry of the linear $D=4$ theory, \fsym.
$Sp(56)$ has a subgroup $GL(27)\times SL(2)$ and the $SL(2) $ in
\hsym\ is a diagonal subgroup of the $SL(2)\times SL(2)$ in $SL(2)\times 
Sp(56)$:
$$GL(27)\times SL(2) \subset GL(27)\times SL(2)\times SL(2)
\subset Sp(56)\times SL(2)
\eqn\abc$$
The 56 vector gauge fields in $D=4$
subject to the constraint \cjeq\
transform as a
 ({\bf 2,27})+({\bf 2,1}) of $ SL(2) \times GL(27)$
while the 70 scalar fields 
transform as a
({\bf 1,27})+({\bf 42,1})+({\bf 1,1}) of $GL(42) \times GL(27)$.
The graviton and its duals transform under $SL(2)$ as in \symten, subject 
to the constraints \rcon.

Thus for the free (4,0) theory, 
reducing on a 2-torus gives a $D=4$ theory with $SL(2)$ symmetry, and this is
indeed a subgroup of the symmetries of the linearised $N=8$ supergravity
theory.
In particular, the 56 vector potentials transforming as a {\bf 56} of
$Sp(56)$ decompose into a 
$({\bf 2,27})+({\bf 2,1})$ under the $ SL(2) \times GL(27)$ subgroup of 
$Sp(56)$.
In the full non-linear supergravity theories,
the interactions break the
$GL(27)$ in $D=5$ to $E_{6}$, with the 27 $D=5$ vector fields transforming as 
a {\bf 27},  and break the $Sp(56)$ in $D=4$ to $E_{7}$, with the 28+28 $D=4$ vector
fields transforming as a
{\bf 56}.
However, although $ SL(2) \times GL(27)$ is a subgroup of $Sp(56)$,
$SL(2)\times E_{6}$ is not a subgroup of $E_{7}$. As $E_{7}$ is the 
maximal symmetry of the vector/scalar sector of the $D=4,N=8$ 
supergravity theory, this $SL(2)$ cannot be a symmetry of the 
supergravity equations of motion, and at most an abelian subgroup 
$\R ^{+}$ can 
survive, as $\R ^{+}\times E_{6}$ is   a maximal subgroup of $E_{7}$.
Moreover, the $SL(2)$ arising from the 
torus reparameterisations necessarily acts on the graviton, while the usual 
U-duality symmetries leave the Einstein frame metric invariant, and 
there is no such symmetry of the non-linear Einstein equations.
This $SL(2)$ symmetry is then not a symmetry of the $D=4,N=8$ 
supergravity field equations.
This raises a number of issues for the interacting theory, which will be 
addressed in the next 
section.

Note that $E_{7}$ does have a maximal subgroup
$SL(2)\times SO(6,6)$ 
which is the product of an $SL(2)$ S-duality and an $SO(6,6)$ 
T-duality under which the
56 vector potentials  consist of the 24 in the NS-NS sector 
transforming as a ({\bf 2,12}), while there are 32 in the RR sector transforming
as a ({\bf 1,32}), so that these are singlets under $SL(2)$ [\HT].
For the $SL(2)$ arising from the reduction of the (4,0) theory on 
$T^{2}$, all the vector fields would be in $SL(2)$ doublets.

\chapter{Symmetries of Interacting Theories}

Consider first the (2,0) theory, compactified on a 2-torus to give a 
$D=4$ theory with 16 supersymmetries and an $SL(2,\Z)$ symmetry 
arising from the large  diffeomorphisms of the torus.
In the free case, the (2,0) theory is well-understood and the 
compactification is straightforward, and the resulting  $D=4$ theory is the
$N=4$ abelian  gauge theory which indeed has an $SL(2,\Z)$ S-duality symmetry 
of the equations of motion.
In the interacting case, the $D=6$ (2,0) theory is only known implicitly
and a natural candidate for the compactified theory is $N=4$ non-abelian  
Yang-Mills theory. However, the $N=4$ Yang-Mills field theory does not have 
an $SL(2,\Z)$ symmetry of the equations of motion, due to the presence 
of explicit vector potentials in the minimal couplings.
There is some evidence that the full theory should have an $SL(2,\Z)$ 
symmetry -- for example, in the holographic description of the $N=4$ 
conformal field theory, this arises from the  $SL(2,\Z)$ U-duality 
symmetry of the type IIB superstring.
The field theory cannot then give a complete 
formulation of the theory. Indeed, as one approaches the conformal 
point in moduli space, the W-bosons become massless and so do the 
magnetic monopoles and the infinite tower of dyons related to the 
W-boson by S-duality [\HTE].
These light states are mutually non-local and there is
no  conventional local field theory that
can  describe all the light degrees of freedom.
At the conformal point, there are no particle states
and one   instead discusses correlation functions of physical 
operators; see e.g. [\intri].
Thus in the interacting case, the fact that there is no $SL(2,\Z)$ 
symmetry of the non-abelian
super-Yang-Mills field equations can be taken not as an argument against 
the symmetry, but as an argument against the formulation as a 
conventional field theory.

The   (4,0) theory is superficially similar.
The free theory in a flat background can be compactified on a 2-torus 
to give linearised $N=8$ supergravity in $D=4$ which indeed has an 
$SL(2,\Z)$ symmetry inherited from the 2-torus diffeomorphisms.
The full non-linear $D=4$ supergravity theory has
local interactions 
which break the $SL(2,\Z)$ duality symmetry; in the gauge theory case,
 S-duality would transform the gauge potential $A$ into its dual
$\ti A$, and the minimal couplings are clearly not invariant, while in 
the linearised gravity theory the graviton $h_{mn}$ is transformed 
into the dual gravitons 
$\hat h_{mn},\ti h_{mn}$ and this symmetry is violated by the 
non-linear couplings to the metric in the full theory.

Either the full $D=4$ theory has such an 
$SL(2,\Z)$ symmetry, or it does not.
If  the full theory does indeed have the $SL(2,\Z)$
symmetry, then the situation would be
similar to the case of the interacting (2,0) 
theory, and the fact that the $N=8$
supergravity theory doesn't have the requisite 
 symmetry would be taken 
not as evidence against the symmetry, but as an indication that the 
supergravity doesn't give a complete description of the theory, just 
as the 
super-Yang-Mills 
theory doesn't give a complete description of the theory 
obtained by reducing the (2,0) theory.

For gravity in any dimension, the  gauge symmetry is
$$\dd g_{\mm\nn}=2\nabla_{(\mm}\xi_{\nn)}
\eqn\dif$$
If the metric is written as
$$g_{\mm\nn}=\bar g_{\mm\nn}+h_{\mm\nn}
\eqn\abc$$
in terms of a fluctuation $h_{\mm\nn}$ about 
  some background metric $\bar g_{\mm\nn}$ (e.g. a flat background  metric)
 then two main types of symmetry emerge. 
 The first consists of
 \lq background reparameterizations'
 $$\dd \bar g_{\mm\nn}= 2\bar \nabla_{(\mm}\xi_{\nn)}
 ,\qq
 \dd
h_{\mm\nn}= {\cal L}_{\xi}h_{\mm\nn}
\eqn\bac$$
where
$\bar \nabla$ is the background covariant derivative 
with connection constructed from $\bar g_{\mm\nn}$, while  
$h_{\mm\nn}$ transforms as a tensor (${\cal L}_{\xi}$ is the Lie 
derivative with respect to the vector field $\xi$),
as do all other covariant fields.
The second is the \lq gauge symmetry'
of the form
 $$\dd \bar g_{\mm\nn}=0, \qq
 \dd
h_{\mm\nn}= 2  \nabla_{(\mm}\zz_{\nn)}
\eqn\gag$$
in which 
$h_{\mm\nn}$  transforms as a gauge field and the background is 
invariant.
There is in addition the standard shift symmetry
under which
$$\dd \bar g_{\mm\nn}= \aa_{\mm\nn}
 ,\qq
 \dd
h_{\mm\nn}=- \aa_{\mm\nn}
\eqn\shift$$
Combining \bac\ with the shift symmetry \shift\ gives \gag\ (or  
forms of the transformation that are intermediate between \bac\ and 
\gag), and in terms of the 
full metric $g_{\mm\nn}$, there is no shift symmetry and a unique gauge
symmetry \dif; the various types of symmetry \bac,\gag,\shift\
are an artifice of the background split.
The shift symmetry is a signal of background independence and plays an
important role in the interacting theory.

The linearised $D=5$ supergravity theory has both 
  background reparameterization and gauge invariances given by the 
  linearised forms of \bac,\gag\ respectively, and both of these have 
  origins in $D=6$ symmetries of the free (4,0) theory.
  The background reparameterization invariance lifts to the 
  linearised $D=6$ background 
  reparameterization invariance
  $$\dd \bar g _{MN}= 2\pa_{(M}\xi_{N)},\qq \dd C_{MNPQ}={\cal L}_{\xi}C_{MNPQ}
  \eqn\abc$$
  with the transformations leaving the flat background metric
$ \bar g _{MN}$ invariant forming the $D=6$ Poincar\' e group.
The $D=5$ gauge symmetry given by the linearised form of \gag\ arises from
the $D=6$ gauge symmetry
$$ \dd C_{MN\, PQ} = \pa _{[M} \chi _{N]PQ} +\pa _{[P} \chi
_{Q]MN}-2
\pa _{[M} \chi _{NPQ]}
\eqn\delciss$$
with $\dd \bar g _{MN}= 0$ and the parameters related by 
$\zz^{\mm}=\chi^{55\mm}$.
The $D=6$ theory has no analogue of the shift symmetry, and the 
emergence of that symmetry on reduction to $D=5$ (and dualising to 
formulate the theory in terms of a graviton $h_{{\mm\nn}}$) comes as a 
surprise from this viewpoint.
On reducing on a 2-torus, the group $SL(2,\Z)$ of large background  
diffeomorphisms
of the 2-torus background give rise to a
symmetry of the resulting $D=4$ theory, just as it would for the 
reduction of any conventional field theory on a 2-torus.

The gravitational interactions of the full supergravity theory in $D=5$ 
are best expressed geometrically in terms of the total metric 
$g_{\mm\nn}$.
If an interacting form of the (4,0) theory exists that reduces to the 
$D=5$ supergravity, it must be of an unusual type.
One possibility is that there is no background metric of any kind in 
$D=6$, and the full theory is formulated in terms of a total field 
corresponding to $C$, with a spacetime metric emerging only in a 
particular background $C$ field and a particular limit corresponding 
to the free theory limit in $D=5$.

It is not even clear that such an interacting theory should be formulated 
in a $D=6$ spacetime. 
In $D=5$, the diffeomorphisms can be taken to act actively, leaving 
spacetime coordinates invariant and transforming the fields, or 
 passively  with the coordinates transforming
as
$$\dd x^{\mm}=\xi^{\mm}\eqn\abc$$
In the (4,0) theory, the parameter $\xi^{\mm}$ lifts to a parameter
$\chi^{MNP}$.
The active diffeomorphisms could
lift to transformations of fields in a $D=6$ spacetime manifold,as in 
\delciss, but 
if
the passive ones were to lift, it could be to something like a 
manifold with coordinates $X^{MNP}$ transforming through 
reparameterisations
$$\dd X^{MNP}=\chi^{MNP}
\eqn\abc$$
with the $D=5$ spacetime arising as a submanifold with
$x^{\mm}=X^{55\mm}$.
If this was the case, the 
picture of a $D=6$ theory compactified on a 2-torus would only emerge 
in the free limit, and the 
lack of an $SL(2,\Z)$ symmetry in the interacting $D=4$ theory would 
reflect the absence of a conventional spacetime picture in the theory 
that emerges in the strong coupling limit of the $D=5$ theory.

Similar considerations apply to the local supersymmetry transformations.
In $D=5$, the local supersymmetry transformations in a supergravity 
background give rise to \lq background 
supersymmetry transformations' with symplectic Majorana spinor  parameters
$\ee ^{\aa a} $ 
(where $\aa$ is a $D=5$  spinor index and $a=1,..,8$ labels the 8 
supersymmetries) in which the 
gravitino fluctuation $\psi_{\mm}^{a}$ transforms without a derivative of
$\ee^{a}$, and 
\lq gauge supersymmetries' with spinor parameter $\varepsilon 
^{\aa a} $
under which 
$$\dd 
\psi_{\mm}^{a}= \pa_{\mm}\varepsilon ^{a} +\ldots
\eqn\ssy$$
The background symmetries preserving a flat space background
form the $D=5$ super-Poincar\' e group.
In the free theory, the $D=5$ super-Poincar\' e symmetry lifts to part 
of a $D=6$ super-Poincar\' e symmetry with $D=5$ translation parameters 
$\xi^{\mm}$ lifting to $D=6$ ones $\Xi^{M}$ and supersymmetry 
parameters $\ee $ lifting to $D=6$ spinor parameters $\hat \ee$.
The corresponding $D=6$ supersymmetry charges $Q$ and momenta $P$
form part of the (4,0) super-Poincar\' e algebra
with
$$
\{Q_\alpha^a,Q_\beta^b\} =\, \www^{ab}\big( \Pi_+
\Gamma^MC\big)_{\alpha\beta} P_M
\eqn\fialgss$$
where $\Pi_\pm$ are the chiral projectors
$$
\Pi_{\pm}=\2 (1\pm \ggg ^{7})
\eqn\abc$$
$\aa,\bb$ are $D=6$ spinor indices and $a,b=1,\ldots,8$ are $USp(8)$ 
indices, with $\www^{ab}$ the
$USp(8)$-invariant anti-symmetric tensor.
This is in turn part of the $D=6$ superconformal group  $OSp^*(8/8)$.

The $D=5$ gravitini lift to 
 fermionic
 2-form
gauge fields $\psi _{MN   }^{\aa a}$ which are  also $D=6$  Weyl spinors ($
\psi _{MN }^\aa=\psi _{[MN] }^\aa
$).
The supersymmetry \ssy\ lifts to the gauge symmetry  
$$\dd \psi _{MN }^a =
\pa_{[M}\varepsilon_{N]}^a+\ldots
\eqn\delyis$$
with parameter a spinor-vector $\varepsilon_{N}^{\aa a}$. 
The   field strength
$$
\chi_{MNP}^{\aa}=\pa_{[P} \psi _{MN] }^\aa
\eqn\abc$$
satisfies
the self-duality constraint
$$
\chi_{MNP}^{\aa a}
= \6 \epsilon _{MNPTUV}
\chi^{TUV\aa a}
\eqn\pdu$$
The $D=5$ gauge symmetries including those with parameters $\zz 
^{\mm},\varepsilon^{a}$ satisfy a local algebra whose global limit is 
the $D=5$ Poincar\' e algebra, but the $D=6$ origin of this (at least 
in the free theory) is an algebra including the
generators ${\cal Q} 
^{a}_{\aa M}$ of the fermionic symmetries with parameter 
$\varepsilon_{N}^{\aa}$ and the generators ${\cal P}^{MNP}$ of the 
bosonic symmetries with parameter $\chi^{MNP}$.
The global algebra is of the form
$$
\{{\cal Q} _{\alpha N}^a,{\cal Q}  _{\beta P}^b\} =\, \www^{ab}\big( \Pi_+
\Gamma^MC\big)_{\alpha\beta} {\cal P}_{(NP)M}
\eqn\fialgsss$$
In the reduction to $D=5$,  the $D=5$ superalgebra has charges
$Q _{\alpha }^a ={\cal Q} _{\alpha 5}^a$, $P_{\mm}={\cal P}_{55\mm}$.

Supersymmetry provides a further argument against the possibility of 
a background metric playing any role in an interacting (4,0) theory 
in $D=6$. 
The $D=5$ supergravity can be formulated in an arbitrary supergravity 
background, but these cannot be lifted to $D=6$ (4,0) backgrounds
involving a background metric
as there is no (4,0) multiplet including a metric or graviton.
The absence of a (4,0) supergravity multiplet appears to rule  out the 
possibility of a background metric  and
the standard supersymmetry \fialgss\
playing any role in the $D=6$ 
theory. Indeed, the interacting theory (if it exists) should 
presumably be a theory based on something like
the algebra \fialgsss\ rather than the
super-Poincar\' e algebra \fialgss.

There seem to be three main possibilities.
The first is that there is no interacting version of the (4,0) theory, 
  that it only exists as a free theory, and that the limit proposed 
  in [\cst] only exists for the free $D=5$ theory.
The second is that an interacting form of the theory does exist in 6 
spacetime dimensions, with $D=6$ diffeomorphism symmetry. The absence of a spacetime metric means that 
such a generally covariant theory must be of  an unusual kind.
If such a theory exists, then for a spacetime with the topology
$T^{2}\times M_{4}$ for some 4-manifold $M_{4}$, then the group $SL(2,\Z)$ 
of large diffeomorphisms of the torus should give rise to  an 
S-duality of the dimensionally reduced theory. 
The $N=8$ supergravity has no such symmetry, so the reduction would 
give not the supergravity but some modification of this theory 
(presumably not both covariant and local) which does have the 
invariance.
This would be similar to the reduction of the interacting (2,0) 
theory, which gives a $D=4$ theory with $SL(2,\Z)$ invariance, and 
which therefore cannot be the $N=4$ super-Yang-Mills field theory.
Such a symmetry would imply that $D=4$ gravity is self-dual, with its 
strong coupling behaviour being governed by an identical theory to 
the weak coupling theory.

The third and perhaps the most interesting 
possibility is that the theory that reduces to the interacting 
supergravities in $D=4,5$ is not a diffeomorphism-invariant theory in 
six spacetime dimensions, but is something more exotic, perhaps of the 
type suggested above. If so, there is no obvious reason to expect an 
$SL(2,\Z)$ symmetry in the full theory,
but it could
  arise as a \lq bonus symmetry' in the free 
limit in which the new theory reduces to the free (4,0) field theory.

\refout

\bye